\documentclass[apjl]{emulateapj}
\usepackage{natbib}
\usepackage{enumitem}
\usepackage{hyperref}
\usepackage{amsmath}
\hypersetup{
    colorlinks=true,
    linkcolor=blue,
    citecolor=blue,
    filecolor=blue,
    urlcolor=blue
}
 \providecommand{\adsurl}[1]{\href{#1}{ADS}}
\usepackage{amsmath,amssymb}
\usepackage{mathptmx}
\usepackage{mathtools}
\DeclareMathAlphabet{\pazocal}{OMS}{zplm}{m}{n}

\usepackage{graphicx}
\defcitealias{Parravano2003}{PHM}
\begin{document}

\title{The far-UV Interstellar Radiation Field in Galactic Disks:
\\ Numerical and Analytic Models}
 \author{Shmuel Bialy$^\star$}
 \affiliation{Harvard Smithsonian Center for Astrophysics, 60 Garden st., Cambridge, MA, 02138
 }
 \email{$^\star$sbialy@cfa.harvard.edu}
\slugcomment{ApJ. Accepted (Sept 10, 2020)}

\begin{abstract}
The intensity of the far-ultraviolet (FUV; 6-13.6 eV) interstellar radiation field (ISRF) in galaxies determines the thermal and chemical evolution of the neutral interstellar gas and is key for interpreting extragalactic observations and for theories of star-formation. We run a series of galactic disk models and derive the FUV ISRF intensity as a function of the dust-to-gas ratio, star-formation rate density, gas density, scale radius, and observer position. We develop an analytic formula for the median FUV ISRF flux. We identify two dimensionless parameters in the problem: (1) the dimensionless galactic radius, $X$, which measures the radial extent of  FUV sources (OB stellar associations) in the disk; (2) the
opacity over the inter-source distance, $\tau_{\star}$,
 which measures the importance of dust absorption.
These parameters encapsulate the dependence on all of the physical parameters. We find that there exists a critical $\tau_{\star}$, or equivalently a critical dust-to-gas ratio, $Z_{d,{\rm crit}}' \approx 0.01-0.1$ the Milky Way value, at which the ISRF changes behavior. For $Z'_d>Z_{d,{\rm crit}}'$ the ISRF is limited by dust absorption. With decreasing $Z'_d$, the ISRF intensity increases as more sources contribute to the flux. For $Z'_d < Z_{d,{\rm crit}}'$ the ISRF saturates as the disk becomes optically thin. We find that the ISRF per star-formation rate density in low metallicity systems, such as dwarf and high redshift galaxies, is higher by up to a factor of 3-6 compared to their Milky-Way counterparts. We discuss implications to the potential mechanisms that regulate star-formation in low metallicity galaxies.

\end{abstract}


\section{Introduction}

Massive stars are responsible for the generation of the 
far-ultraviolet (FUV; $6-13.6$ eV) interstellar radiation field (ISRF) in galaxies.
The intensity of the FUV ISRF
 is a key property of the interstellar medium (ISM) of galaxies. 
 It shapes the thermal and chemical structure of the ISM and may be key for regulating the star-formation rate (SFR).
The FUV radiation is typically the dominant heating source of the neutral ISM, through the photoelectric ejection of electron off dust grains \citep{Bakes1994}, and it thus controls
the thermal state of interstellar gas and the balance between the warm-cold neutral media (WNM/CNM) phases \citep{Wolfire1995, Liszt2002, Wolfire2003, Hill2018, Bialy2019}.
The FUV photoelectric heating introduces a natural feedback loop for star formation in galactic disks,  where any increase in the star-formation rate (SFR) results in an increased gas heating rate which reduces the abundance of the cold (and dense) phase reducing back the SFR \citep{Corbelli1988, Parravano1988,  Parravano1989, Ostriker2010}.

Subsequent studies 
\citep{Kim2011, Ostriker2011, Shetty2012, Kim2013b, Kim2015, Kim2017},
stressed the importance of supernova (SN) feedback as a regulation mechanism 
for star-formation. 
These studies find that momentum injection by SNe generates turbulent pressure that exceeds the thermal pressure, suggesting that SN feedback, rather than FUV photoelectric heating, is the dominant driver of star-formation.
Interestingly, we find that in low metallicity systems (such as dwarfs and high redshift galaxies)
the ratio of the FUV ISRF per SFR is elevated by a factor of 3-6 (compared to Milky-Way-like galaxies), implying that the thermal pressure in these systems may  become important and that the FUV photoelectric heating may be the dominant self-regulation process of star-formation (see \S \ref{sub: SF regulation} for an elaborate discussion).
A key factor in the self-regulation theory is the ratio of the FUV ISRF intensity to the SFR density.
While often assumed to be constant,
we show that this ratio varies with galaxy properties: the gas density, SFR, dust-to-gas ratio (DGR), and galactic scale radius.
In this paper we characterize this dependence through a set of numerical calculations, and an analytic model.

The intensity of the FUV ISRF also determines the chemical evolution of atomic and molecular clouds in the ISM, and the intensity of molecular radio and sub-millimeter emission lines through photo-excitation and photo-dissociation of various molecular energy levels at the edges of photodissociation regions \citep[PDRs][]{Tielens1985, Hollenbach1999}.
A model of the FUV ISRF is thus required for an accurate interpretation of extragalactic observations tracing the atomic/molecular ISM.
In particular, the FUV intensity is a key parameter for the process of atomic-to-molecular (HI-to-H$_2$) transition \citep{Glassgold1974, Draine1996, Krumholz2008, McKee2010, Sternberg2014, Bialy2016a, Bialy2017}, which may regulate and/or trace star-formation \citep[e.g.,][]{Leroy2008, Bigiel2008, Gnedin2009, Glover2012, Kuhlen2012, Diemer2019a}.

The FUV ISRF is the summed contribution of FUV fluxes from OB stellar associations, down-weighted by the effect of dust absorption \citep[][hereafter \citetalias{Parravano2003}]{Parravano2003}.
 As such, the FUV ISRF is a function of various physical parameters that characterize the galaxy:
 the dust-to-gas ratio (DGR) or the galaxy metallicity, the gas mean density, and the density of OB stellar associations which is in turn related to the SFR.
The aim of this paper is to develop an analytic model to describe the FUV ISRF intensity
as a function of these physical properties.

 The structure of the paper is as follows.
We describe the model ingredients and our numerical calculations in \S \ref{sec: numerical model}. In this section we also identify the governing dimensionless parameters and connect them to the physical parameters, the SFR density, the gas density, the DGR and the OB stars scale radius. 
In \S \ref{sec: results} we present the numerical results and derive an
an analytic formula for the FUV ISRF.
We do so in increasing steps of complexity, starting from an idealized model of an observer located in the disk center, and FUV sources of constant luminosity (\S \ref{sub: th const lum}).
We then generalize the theory to: 1) account for a source luminosity distribution that resembles that of realistic OB stellar associations (\S \ref{sub: th luminosity distrbution}), and 2) to describe off-central observers (e.g., the sun in the Milky-Way galaxy) where the polar symmetry of the problem is broken (\S \ref{sub th non center}).
We summarize our model and discuss implications to star-formation self-regulation and potential model extensions in \S \ref{sec: discussion}. We conclude in \S \ref{sec: conclusions}.

\section{Model Ingredients}
\label{sec: numerical model}


Here we discuss the basic model ingredients for the derivation of the FUV ISRF for galaxies of different properties: galaxy radius, DGR, and SFR.
We assume a thin exponential disk, in which we randomly distribute radiation sources (see \S \ref{sub: numerical procedure} for more details), where the radiation sources represent associations of OB stars.
At a given point in the disk,  $\vec{r}$,  the FUV ISRF is the sum of all FUV radiation sources, and is given by
\begin{equation}
\label{eq: F basic}
   F(\vec{r}) = \sum_{i} \frac{L_i \textrm{e}^{-\sigma n | \vec{r}_i-\vec{r}| }}{4 \pi |\vec{r}_i-\vec{r}|^2} \ ,
\end{equation}
where $\vec{r}_i$ and $L_{i}$ are the position and luminosity of each source.
In the exponent, $\sigma$ is the dust absorption cross-section per hydrogen nucleus (over the FUV band),  and $n$ is the mean hydrogen nucleus density of the gas.
Assuming an exponential disk, the surface density of FUV sources 
decreases as
\begin{equation}
\label{eq: N(r)}
    N_{\star}(r) = N_{{\star},c} \ \mathrm{e}^{-r/R_{\rm gal}} \ ,
\end{equation}
where $N_{\star c}$ is the central surface density, and $R_{\rm gal}$ is the exponential scale-radius of OB associations.

Since OB associations vary in mass and luminosity, there is no single value for the source luminosity, $L_i$. 
We adopt the probability density function (PDF)
\begin{align}
\label{eq: dp/dL, dp/dlnL}
    \frac{\mathrm{d}p}{\mathrm{d}L} = \frac{1}{\Lambda-1} \frac{L_{\rm max}}{L^2}  
\end{align}
when $L_{\rm min} \leq L \leq L_{\rm max}$, and $\mathrm{d}p/\mathrm{d}L=0$ otherwise.
Here $L_{\rm min}$, $L_{\rm max}$ are the luminosities of the least and most luminous associations, and
\begin{equation}
    \Lambda \equiv L_{\rm max}/L_{\rm min}  = 5900 \ .
\end{equation}
This source luminosity distribution is derived from the distribution of number of stars in OB associations discussed in \citetalias{Parravano2003}\footnote{We obtain the cumulative distribution for luminosity larger than $L$ by following Eq.~15 in \citetalias{Parravano2003}, and expressing the distribution in terms of association luminosity  assuming it is proportional to the number of massive stars in it.}, based on 
the \citet{McKee1997} model.
The mean luminosity per source is $\langle L \rangle= \ln \Lambda/(\Lambda-1) L_{\rm max}=8.7 L_{\rm min} = 1.5\times 10^{-3} L_{\rm max}$.


\subsection{Governing Dimensionless parameters}
\label{sub: dimensionless units}
In this subsection we identify the dimensional and
dimensionless quantities. We use them to rewrite the problem in dimensionless form (Eq.~\ref{eq: f_basic} below). This makes our results more general and allow for future variations and extensions to our model (see \S \ref{sub: advantage of dimensionless form} for a discussion).

The natural flux unit is the source emissivity
\begin{equation}
    \Sigma_{\star} \equiv \langle L \rangle N_{\star}
\end{equation}
where $N$ is the surface density of all FUV sources (i.e., OB associations of any luminosity) at the position of the observer.
There are three distance scales that naturally appear in the problem: (a) the inter-source scale
\begin{equation}
    l_{\star} \equiv N_{\star}^{-1/2} \ ,
\end{equation}
(b) the dust-absorption length-scale 
\begin{equation}
    R_d \equiv 1/(\sigma n) \ ,
\end{equation}
and (c) the galactic scale radius, $R_{\rm gal}$.
These length-scales may be combined to give two independent dimensionless parameters, which characterize the problem:
\begin{align}
    \tau_{\star} &\equiv l_{\star}/R_{\rm d} = \sigma n l_{\star}
    \\ \nonumber
        \tau_{\rm gal} &\equiv R_{\rm gal}/R_{\rm d} = \sigma n R_{\rm gal}
\end{align}
These parameters are 
the dust opacity over the inter-source distance, and the dust opacity over the galactic scale radius, respectively.
Alternatively, we will often use the combination of the parameters:
$\tau_{\star}$ and the dimensionless galactic radius,
\begin{equation}
    X \equiv R_{\rm gal}/l_{\star} = \tau_{\rm gal}/\tau_{\star} \ .
\end{equation}
$X$ measures the extent of OB stellar associations in the galactic disk, and is independent of the dust abundance.

  \begin{figure*}[t]
	\centering
		\includegraphics[width=1\textwidth]{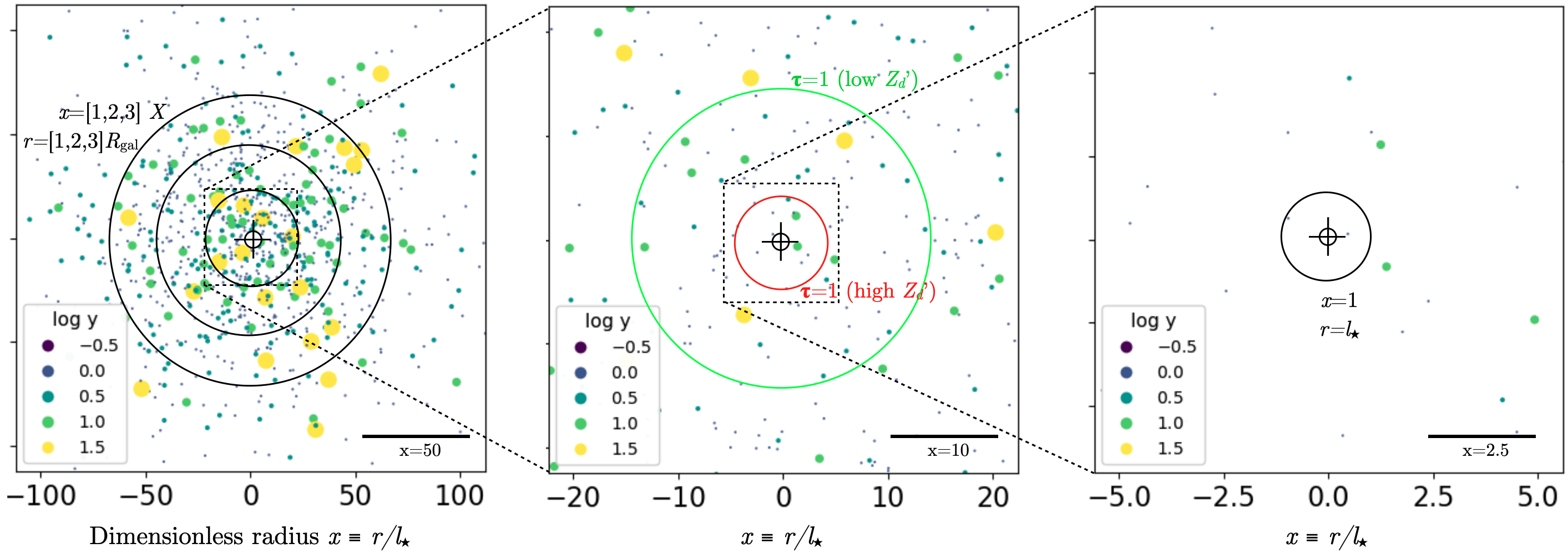} 
	\caption{
    One random realization (out of 20k for each model) of OB associations in an exponential disk as given by our numerical model.
    The $x-y$ coordinates are in units of the inter-source distance $l_{\star}$ (see Eq.~\ref{eq: f_basic}).
    The colored markers (also vary in size) show the associations binned by $\log_{10} y$ where $y\equiv L/\langle L \rangle$ following Eq.~(\ref{eq: dp/dy}). 
    Also shown are the galactic scale radii $r= [1,2,3]R_{\rm gal}$ (i.e., $x=[1,2,3]X$) (left panel), the dust-absorption radius where $\tau=1$ (i.e., $r=R_d$ or $x=\tau_{\star}^{-1}$) (middle-panel), and the  inter-source scale $r=l_{\star}$ (or $x=1$) (right-panel).
	}
		\label{fig: 2d sources}
\end{figure*}

\subsection{Connection to Physical Units}
\label{sub: physical units}

The values of $\tau_{\star}$ and $X$ (and $\tau_{\rm gal}$) 
depend on the DGR, gas density, and the SFR density.
These parameters may vary for galaxies of different masses, evolutionary stages, and with cosmic epoch, as well as with galactocentric radius, for a given galaxy.

Assuming that the surface density of OB associations is proportional to the SFR surface density we write
\begin{equation}
\label{eq: N sfr}
    N_{\star} = 6.8 \times 10^{-5} \ \Sigma_{\rm sfr}^{\prime} \ {\rm pc}^{-2} \ ,
\end{equation}
where $\Sigma'_{\rm sfr}$ is the SFR surface density normalized to solar-circle, 
and the normalization $N_{{\star}\odot}=6.8 \times 10^{-5} \ {\rm pc^{-2}}$ is the total density of OB associations at the solar circle (\citetalias{Parravano2003}).
We adopt a mean source luminosity $\langle L \rangle =1.8 \times 10^{5} \ L_{\odot}$,
giving a source emissivity
\begin{equation}
\label{eq: Sigma sfr}
    \Sigma_{\star} = N_{\star} \langle L \rangle = 12 \ \Sigma_{\rm sfr}^{\prime} \ L_{\odot} \ {\rm pc}^{-2} \ ,
\end{equation}
in good agreement with \citetalias{Parravano2003} 
 who obtain $\Sigma_{{\star}}=10.9 \ L_{\odot} \ {\rm pc}^{-2}$ for the solar circle.
With these assumed values for $N_{\star \odot}$ and $\langle L \rangle$, our numerical calculations recover the observed FUV ISRF,
 $F_{\odot}=2.7 \times 10^{-3}$ erg cm$^{-2}$ s$^{-1}$ 
 \citep{Draine1978}, for solar neighborhood conditions. 

For a given SFR density, gas density, and DGR, 
\begin{align}
\label{eq: l_star physical}
    l_{\star} &= N_{\star}^{-1/2} = 121 \ \Sigma_{\rm sfr}'^{-1/2}  \ {\rm pc} \\ 
    \nonumber
    R_d &= 1/(\sigma n) = 324 (n_0 Z_d')^{-1} \ {\rm pc} \ , 
\end{align}
where we defined $n_0 \equiv n/({\rm cm^{-3}})$, and assumed $\sigma=10^{-21} Z_d'$ cm$^2$ where
$Z_d'$ is the DGR normalized to the solar circle value.
The value of $l_{\star}$ represents a characteristic distance between OB associations (of any luminosity) \footnote{Since the probability per association strongly decreases with $L$, $N_{\star}$ is dominated by under-luminous associations (compared to $\langle L \rangle$), and 
$l_{\star}$ is also a typical distance for these associations.
Super-luminous associations are more rare and have a larger $l_{\star}$ value.}.
The dimensionless parameters
\begin{align}
\label{eq: l_star tau_star to physical}
    \tau_{\star} &= \sigma n l_{\star} = 0.37 \ n_0 \ Z_d' \  \Sigma_{\rm sfr}'^{-1/2} \\ \nonumber
    \tau_{\rm gal} &= \sigma n R_{\rm gal} = 3.1 \ n_0 \ Z_d' \ R_{\rm gal, kpc}
    \\ \nonumber
    X &= R_{\rm gal}/l_{\star} = 8.2 R_{\rm gal, kpc} \Sigma_{\rm sfr}'^{1/2}
\end{align}
where $R_{\rm gal, kpc} \equiv R_{\rm gal}/{\rm kpc}$.
For solar neighbourhood conditions $n_0 \approx 1$, $\Sigma_{\rm sfr}' = Z_d'=1$ (by definition),
and we get
$l_{\star  {\odot}} \approx 121$ pc, and $\tau_{\star, {\odot}}\approx 0.37$.
Assuming an OB galactic scale radius $R_{\rm gal}=3.5$ kpc \citep{Wolfire2003}, we further obtain $X_{\odot} \approx 29$ for the solar-circle.
The values of the parameters above depend on the SFR density, gas density, DGR and galactic scale radius. 
With increasing gas density and/or DGR, the effective dust-absorption scale, $R_d$, decreases and the opacities, $\tau_{\star}$ and $\tau_{\rm gal}$, rise.
As $\Sigma_{\rm sfr}'$ increases the density of sources increases, leading to a lower $l_{\star}$ and $\tau_{\star}$, and a higher $X$.
Finally  $X$ also increases with the galactic scale radius of OB associations.

While $\tau_{\star}$ and $X$ are expected to vary for different galaxies, 
they also vary within a galaxy, with varying galactocentric radius of the observer, $R_{\rm obs}$.
In Eq.~(\ref{eq: l_star tau_star to physical}) this dependence 
is encapsulated in the SFR density parameter, where $\Sigma_{\rm sfr}' \propto N_{\star} \propto \mathrm{e}^{-R_{\rm obs}/R_{\rm gal}}$.
Defining the dimensionless observer radius
\begin{equation}
\label{eq: delta_def}
    \delta \equiv R_{\rm obs}/R_{\rm gal}
\end{equation}
the values of $l_{\star}$, $\tau_{\star}$ and $X$ at the observer position are related to the disk center values through
\begin{align}
\label{eq: local to central}
l_{\star}/l_{\star c} &= \tau_{\star}/\tau_{\star c} = \mathrm{e}^{+0.5 \delta} \\ \nonumber
X/X_c &= \mathrm{e}^{-0.5 \delta} \ ,
\end{align}
where $l_{\star c}$, $\tau_{\star c}$, and $X_c$ are the parameter values at disk center\footnote{Throughout the paper when we include a subscript $c$ it means the parameters are evaluated at disk center. Otherwise, they are evaluated at the observer position.}.
These scalings are necessarily idealized. Realistic galaxies may have regions that do not follow Eq.~(\ref{eq: N(r)}) (e.g., the central region in the Milky-Way), as well as include additional dependence on $\delta$ due to variations in the DGR and gas density. 
Nevertheless, we shall adopt this exponential model as it allows the derivation of generic scaling relations across a large range of physical conditions.

\subsection{Numerical Method}
\label{sub: numerical procedure}
We calculate the FUV ISRF numerically for a large set of galaxy disk models, considering different realization of the radiation sources and placing the observer in various locations within the disk.
To this end we solve the dimensionless form of Eq.~(\ref{eq: F basic}).
Defining the dimensionless: flux $f = F/\Sigma_{\star c}$, distance
$x\equiv r/l_{\star c}$, and source luminosity $y \equiv L/\langle L \rangle$,
Eq.~(\ref{eq: F basic}) takes the form
\begin{equation}
\label{eq: f_basic}
    f(\vec{x}) = \sum_{i} \frac{y_i \textrm{e}^{-\tau_{\star c} |\vec{x}-\vec{x}_i|}}{4 \pi |\vec{x}-\vec{x}_i|^2} \ .
\end{equation}
where the subscripts $c$ refer to the parameters at disk center.
The dimensionless source luminosity follow the PDF 
\begin{align}
\label{eq: dp/dy}
    \frac{\mathrm{d}p}{\mathrm{d}y} = \frac{1}{\ln \Lambda} \frac{1}{y^2}   \ ,
\end{align}
for $y_{\rm min} \leq y \leq y_{\rm max}$ and 0 otherwise, where
 $y_{\rm min}=(\Lambda-1)/(\Lambda \ln \Lambda)$ and $y_{\rm max}=(\Lambda-1)/(\ln \Lambda)$.
 By definition, $\langle y \rangle=1$.
 In these dimensionless units, 
 the distance between sources (at the disk center) is of order unity,
 the galactic scale radius is $X_c$,
 and the total number of sources is $2 \pi X_c^2$.
Thus, each galaxy disk model is characterized by two dimensionless numbers: $\tau_{\star c}$ and $X_{c}$.
While $\tau_{\star c}$ and $X_c$ totally define each of the galaxy disk models, $f$ also depends on the observer position $\delta$.

We consider $13 \times 7$ models, corresponding to all $(\tau_{\star c},  
X_{c})$ pairs within the arrays
$\log_{10} (\tau_{\star c}/\tau_{\star c,0}) =[-2.5, -2.25,..., 0.5]$ and $\log_{10} (X_{c}/X_{c,0})=[-1, -0.75,... 0.5]$ where $\tau_{\star c,0}=0.11$, $X_{c,0}=96$ is our fiducial model for which the solar-circle conditions are recovered at $R_{\rm obs}=8.5$ kpc (see \S \ref{sub: physical units}).
These 91 models represent galaxies of various DGR, gas density, SFRs and galactic radii.
For each model we follow the steps:
\begin{enumerate}
    \item We randomly distribute sources within a two-dimensional disk with 
    an exponentially decreasing source density (Eq.~\ref{eq: N(r)}).
    The source luminosities are randomly drawn from the luminosity PDF (Eq.~(\ref{eq: dp/dy}).
\item We calculate $f$ numerically following Eq.~(\ref{eq: f_basic}), considering 7 different observer positions:
$\delta=0, 0.5, 1, 1.5, 2, 2.4, 3$ ($\delta=2.4$ corresponds to the solar circle).
\item We repeat steps (1-2), 20,000 times to obtain many realizations of source luminosity and positions in the disk. 
These realizations capture fluctuations in the source distribution as new OB associations continuously form while old ones decay.
From the 20k realizations we obtain the distribution of $f$ 
and calculate the median and interquartile range (IQR) for each galaxy model.
\item 
We translate $f$ to  physical units via 
$F=\Sigma_{\star c} f$ (Eq.~\ref{eq: Sigma sfr}).
We derive the normalized ISRF-to-SFR ratio
$I_{\rm UV}/\Sigma_{\rm sfr}'$ where $I_{\rm UV} \equiv F/F_{\odot}$ is the ISRF in units of the solar neighbourhood value.
\end{enumerate}

In Fig.~\ref{fig: 2d sources} we show an example of a single random realization of FUV sources in an exponential disk model with $X=21$, with various levels of zoom-in.
The relative source luminosities, $y\equiv L/\langle L \rangle$, are drawn from the $y$-PDF (Eq.~\ref{eq: dp/dy}), as indicated by the marker-color and marker-size binning (see legend).
In the left-panel (large-scale view) we mark the exponential decline of the source density by showing three circles that correspond to $x=[1,2,3]X$ (or equivalently, $r=[1,2,3]R_{\rm gal}$). 
The value of $\tau_{\star}$ determines the ``radius of influence" within which radiation may travel freely from the sources to the observer (for $x \gg \tau_{\star}^{-1}$ the fluxes vanish due to dust-absorption).
Depending on the value of $\tau_{\star}$ this radius may be small or large. This is shown in the middle panel.
Finally, in the right-panel we show a closeup view.
Here we mark the inter source distance, $x=1$ (or $r=l_{\star}$), which resembles the typical FUV source separation distance.

In addition to the models with the realistic source luminosity distribution, we also run models of constant source luminosity, $y_i=1$, which we use to test our analytic model in the next section.

\section{Results}
\label{sec: results}
In this section we 
present our numerical results for the median FUV ISRF, and the IQR (25-75 percentile range), as functions of galactic properties as 
captured by the $\tau_{\star}$ and $X$ parameters, for various observer positions.
We also derive an analytic model for the median ISRF.
We start with the basic model which assumes constant luminosity sources, and a central observer (\S \ref{sub: th const lum}), and then we generalize the model to 
the case of a source luminosity distribution, and off-disk-center observers (\S\S \ref{sub: th luminosity distrbution},  \ref{sub th non center}).



\subsection{Basic analytic Model}
\label{sub: th const lum}

To obtain an analytic solution for the ISRF we need to make simplifying assumptions:
We assume that the observer is located at disk center. The problem then obeys polar symmetry.
We assume the sources all have the same luminosity (i.e., $y_i=1$ in Eq.~\ref{eq: f_basic}).
Finally, we approximate Eq.~(\ref{eq: f_basic}) as the sum of two components, the contribution of the nearest source plus an integral over the contribution of the rest of the sources.
We get
\begin{align}
\label{eq: f_th}
    \frac{F}{\Sigma_{\star}} 
    &= \frac{1}{4 \pi x_0^2} \mathrm{e}^{-x_0 \tau_{\star}} + \int_0^{2 \pi} \int_{x_0}^{\infty} 
    \frac{1}{4 \pi x} \mathrm{e}^{-\tau_{\star} x} \mathrm{e}^{-x/X} \ \mathrm{d}x \mathrm{d}\theta \\ \nonumber
    &= 
    \frac{1}{4 \pi x_0^2}\mathrm{e}^{-x_0\tau_{\star}} + \frac{1}{2} E_1 \left( x_0 \tau_{\star} + \frac{x_0}{X} \right) \ ,
\end{align}
where $E_1$ is the first exponential function.
The parameter $x_{0}$ is the typical (median) distance at which the nearest source is found.
For randomly placed sources, the number of sources within a given area follows the Poison distribution. 
The area, $a_0$, within which the median number of sources is 1, is
obtained from the solution to the quadratic equation $(N_{\star} a_0)^2 - (2/3) (N_{\star} a_0) - 0.02 = 0$,
giving $N_{\star} a_0= 0.695$.
The corresponding dimensionless radius is
\begin{equation}
\label{eq: r1}
    x_0 = \sqrt{a_0 N_{\star}/\pi} \approx 0.47 \ .
\end{equation}
In Eq.~(\ref{eq: f_th}) we deliberately omitted the subscripts, $c$ (indicating the disk center). While in our derivation we assumed a central observer, as we discuss below, Eq.~(\ref{eq: f_th}) may be also applied to off-center observers under a simple transformation (see \S \ref{sub th non center})


The two exponential factors in Eq.~(\ref{eq: f_th}) account for the exponential attenuation due to dust absorption, and the decrease of the source density with galactocentric radius.
The relative importance of these factors introduces two important limiting cases for Eq.~(\ref{eq: f_th}):
\begin{enumerate}
    \item The weak dust-absorption regime: $\tau_{\rm gal} = \tau_{\star}X \ll 1$
\item The strong dust-absorption regime: $ \tau_{\rm gal} = \tau_{\star}X \gg 1$
\end{enumerate}
The critical point that defines the transition for the two regimes is $\tau_{\rm gal}=1$, or
\begin{equation}
\label{eq: tau_star_crit}
    \tau_{\rm \star, crit}=1/X \ .
\end{equation}
This corresponds to the point at which the dust-scale $R_d=1/(\sigma n)$ is comparable to the galaxy scale radius $R_{\rm gal}$.
This occurs at
\begin{equation}
\label{eq: Zd crit}
    Z_{d,{\rm crit}}' = 0.32 (n_0 R_{\rm gal, kpc})^{-1} \ .
\end{equation}
I.e., at a DGR $\approx 10$\% the Milky-Way value (for $R_{\rm gal, kpc}=3.5$).
We shall now discuss these two limiting cases.

\subsubsection{The weak dust-absorption limit}
In this limit $\tau_{\rm gal}\ll 1$ and dust absorption is negligible over galactic scales. 
The ISRF depends only on $X$ and is independent of $\tau_{\star}$, as the flux is limited by the galaxy scale-radius, i.e., by the exponential decline of the density of FUV sources.
This limit applies to low metallicity and/or small galaxies, e.g., dwarf galaxies and some high redshift galaxies.
	
In this limit, the first term inside the argument of $E_1$ is negligible and we get $F/\Sigma_{\star} \simeq 0.36 + 0.5 E_1[0.47/X]$ 
(since $\tau_{\star} \ll \tau_{\rm gal}$, the exponent in the first term is $\approx 1$).
The contribution of the nearby source (first term) is typically small compared to the summed contribution of all the rest of the sources (the integral).
Their ratio is $0.72/E_1(0.47/X) = 4.6$ to $8.1$ \%, where the numerical values correspond to $X = 300$ to 30. For large $X$ we may also expand $E_1$. 
To leading order $F/\Sigma_{\star} \propto \ln(X)$.
This logarithmic dependence reflects the fact that within $l_{\star} < r < R_{\rm gal}$, the contribution per radial ring falls
off as $1/r$, as the flux from each source is $\propto r^{-2}$ while the number of sources within $(r,r+\mathrm{d}r)$ is $\propto r$.
Indeed, if instead of an exponential disk we assume a constant density disk with a finite radius $R_{\rm gal}$ the integral is $\propto \int (1/r) dr = \ln(R_{\rm gal}/l_{\star})$, where $l_{\star}$ and $R_{\rm gal}$ are the minimum and maximum distances to the radiation sources.

\subsubsection{The Strong dust-absorption limit}
In this limit $\tau_{\rm gal}\gg 1$ and the ISRF is independent of $X$ and depends only on $\tau_{\star}$, as the flux is limited by dust absorption. 
In other words, the characteristic distance at which dust-absorption becomes important, $R_d=1/(\sigma n)$, is much smaller than the galactic scale radius $R_{\rm gal}$.
This is the limit that applies to the Milky-Way, for which $Z_d' > Z'_{d,{\rm crit}}$ (Eq.~\ref{eq: Zd crit}).

In this limit $F/\Sigma_{\star} \simeq 0.36 \mathrm{e}^{-0.47 \tau_{\star}} + 0.5 E_1[0.47 \tau_{\star}]$.
The relative contribution of the nearest source may become substantial, especially at large $\tau_{\star}$.
For $\tau_{\star}$ ranging from 0.05 to 0.5 the ratio of the first to second term ranges from 8.9 to 24 \%.
The $E_1$ function may be expanded if $\tau_{\star}$ is small.
To leading order, the integral is $\propto \ln(1/\tau_{\star})=\ln(R_d/l_{\star})$.
This is similar to what one gets for a constant density disk of radius $R_d$.
Thus, in the strong-dust absorption limit, dust absorption plays the role of an effective galactic edge.

\subsection{Generalization to a Source luminosity distribution}
\label{sub: th luminosity distrbution}

\begin{figure}[t]
	\centering
		\includegraphics[width=0.5\textwidth]{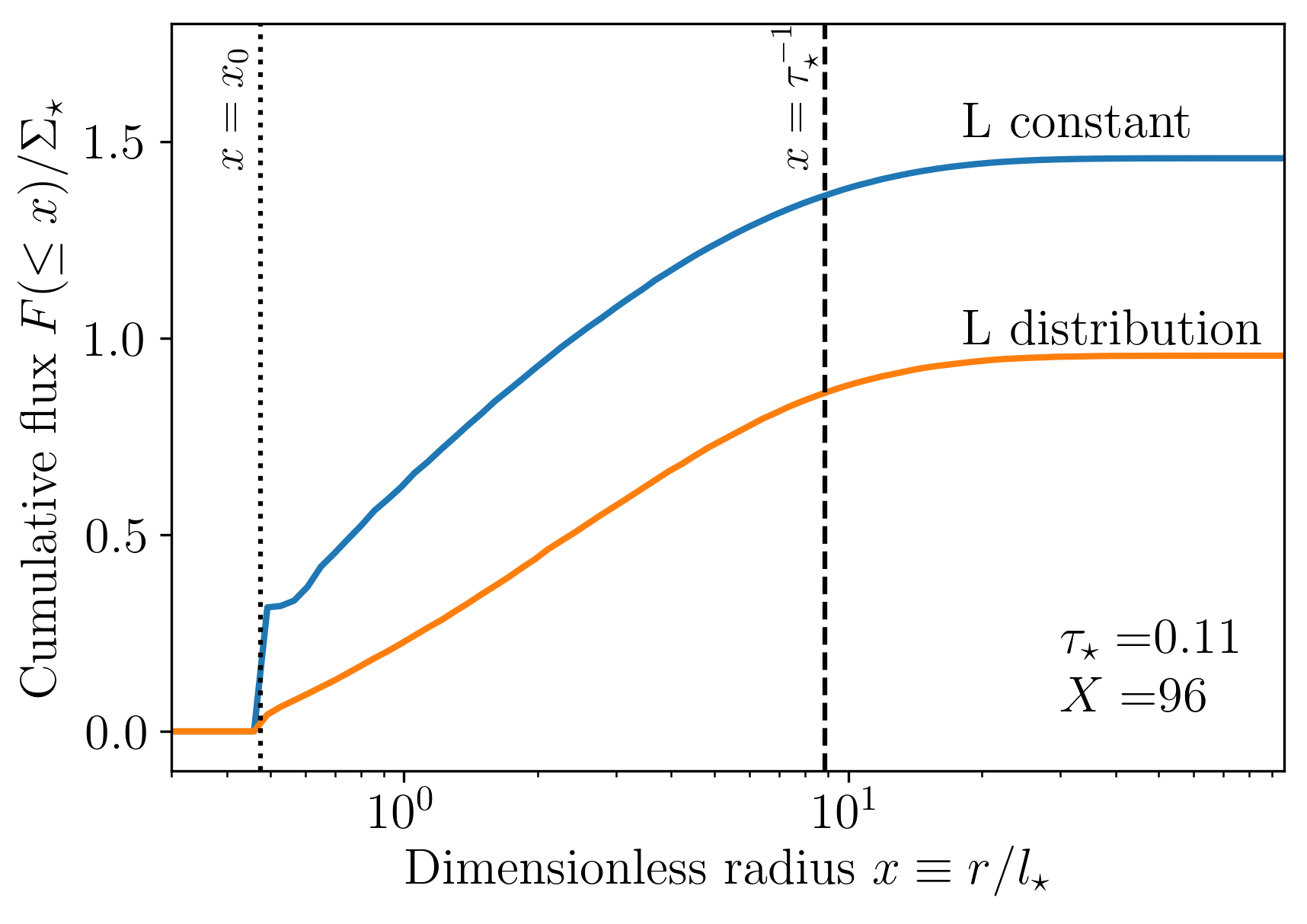} 
	\caption{
	The cumulative flux (median) as a functions of dimensionless distance from the source $x \equiv r/l_{\star}=\tau/\tau_{\star}$, for our fiducial Milky-Way model  ($\tau_{\star}=0.11, X=96$).
	The two vertical lines show the location where the nearest source is encountered ($x_0=0.47$), and the point where the optical depth approaches unity ($x=1/\tau_{\star}$), past which the ISRF saturates.
	The two curves correspond to a model with constant luminosity FUV sources, and to a a model where the sources are described by a luminosity PDF (Eq.~\ref{eq: dp/dy}), representing the luminosity of OB stellar associations.
	While in the constant luminosity source model there is a jump in the ISRF at $x=x_0$ due to the contribution of the nearby source, for a source luminosity distribution the nearby FUV source typically has a low luminosity and thus the increase is gradual.
	}
		\label{fig: cummulative}
\end{figure}

\begin{figure}[t]
    \centering
    \includegraphics[width=0.5\textwidth]{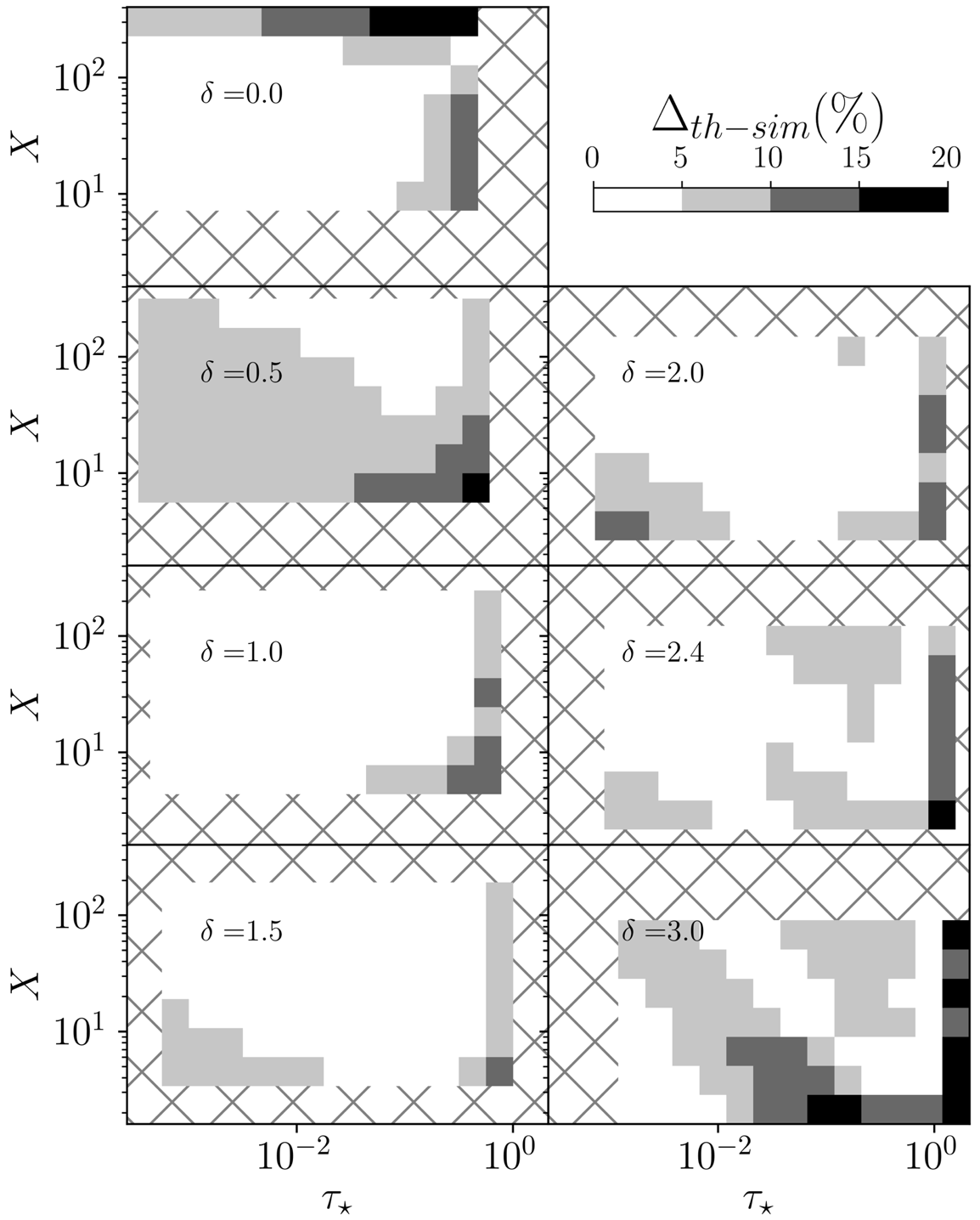}
    \caption{Comparison of our extended analytic model (Eq.~\ref{eq: IUV/SFR}, \ref{eq: X_delta}) with our numerical results. We show the relative difference in $I_{\rm UV}/\Sigma_{\rm sfr}'$ ($\|$Theory-Simulations$\|$/Simulations)  as a function of $\tau_{\star}, X$ and observer position $\delta$.
    The hatched regions are regions outside the numerical model parameter space.
    } 
    \label{fig: comparison}
\end{figure}

When considering a luminosity distribution for the sources,
we can no longer approximate Eq.~(\ref{eq: F basic}) as a sum of two contributions as we did in Eq.~(\ref{eq: f_th}).
This is because the probability per unit area to encounter a source depends on its luminosity (e.g.,  higher luminosity associations are less likely; Eq.~\ref{eq: dp/dy}), and thus $x_0$ becomes a function of $y$.
An approximate analytic expression may be derived
by using an effective value
\begin{equation}
\label{eq: x0 eff}
    x_0 \rightarrow \bar{x}_0 = 0.90 \ ,
\end{equation}
in Eq.~(\ref{eq: f_th}).
This effective value is larger than $x_0$ because the nearest sources are typically of very low luminosity ($y_i \ll 1$), and a larger distance is required to achieve an appreciable contribution to the flux.
This is illustrated in Fig.~\ref{fig: cummulative}, where we plot the calculated cumulative median flux as a function of the source distance, for the case of constant luminosity sources and a luminosity distribution.
For constant luminosity sources, $F/\Sigma_{\star}$ experiences a significant jump at $x=x_0$ from 0 to $\mathrm{e}^{-\tau_{\star}x_0}/[4 \pi x_0^2]$  as the first source is encountered.
On the other hand, in the case of a source luminosity distribution, the first source has a typical luminosity $y\approx y_{\rm min}\ll 1$, and thus it induces only a very mild jump in the flux (barely visible in the figure) and the flux builds up gradually with increasing $x$.
The effective distance $\bar{x}_0$ reflects this smoothing effect.
Comparing the numerical and analytic results we find that 
over most of our parameter space, the agreement is within $\lesssim 10\%$ (see Fig.~\ref{fig: comparison}).

Fig.~\ref{fig: cummulative} also shows that the cumulative fluxes saturate shortly after the distance $x=1/\tau_{\star}$ is reached. 
This corresponds to the condition that the dust optical depth $\tau=1$.
This is in agreement with the strong dust absorption limit, where dust limits the integral of the flux. 
Indeed the displayed model has $\tau_{\star}=0.11$ and $X=96$ so that $\tau_{\star} \gg \tau_{\rm \star, crit}$, corresponding to the strong-dust absorption regime.

\begin{figure}[t]
	\centering
	\includegraphics[width=0.5\textwidth]{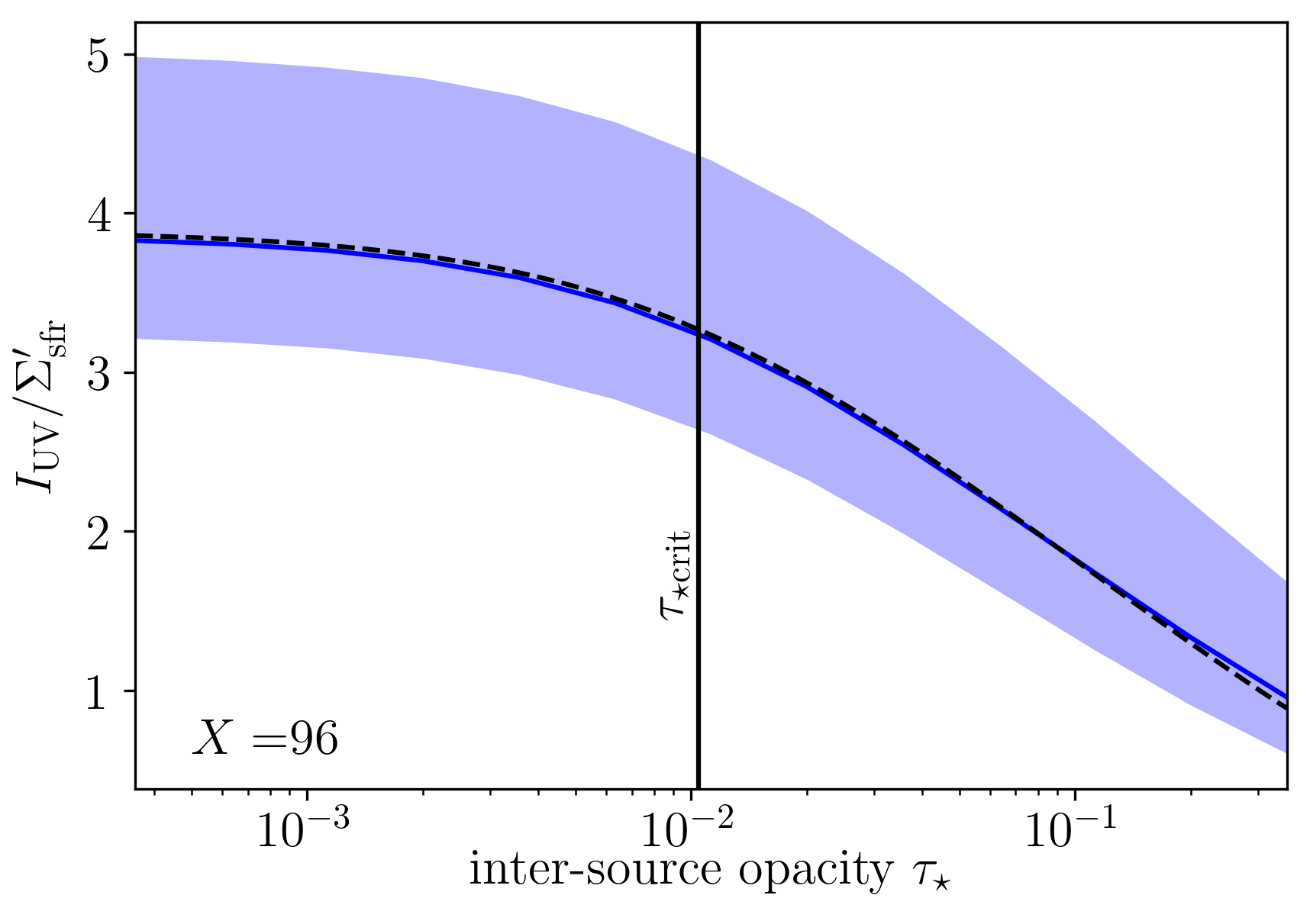} 
	\caption{
	The FUV ISRF intensity to SFR density ratio (normalized to solar circle values) as a function of the inter-source dust opacity $\tau_{\star}$ assuming $X \equiv R_{\rm gal}/l_{\star}=96$ and a central observer.
	The solid curve and the shaded region are the median 
	 and the IQR
	as obtained by our numerical simulation. The dashed curve is the analytic model
	(Eq.~\ref{eq: x0 eff}, \ref{eq: IUV/SFR}).
	The vertical line shows the critical $\tau_{\rm \star crit}$ (Eq.~\ref{eq: tau_star_crit}) above which
    the ISRF limited by dust absorption. For  $\tau_{\star} \ll \tau_{\rm \star crit}$ the ISRF saturates and becomes independent of $\tau_{\star}$.
		}
		\label{fig: IUV_vs_Zd}
\end{figure}

\begin{figure*}[t]
	\centering
	\includegraphics[width=0.85\textwidth]{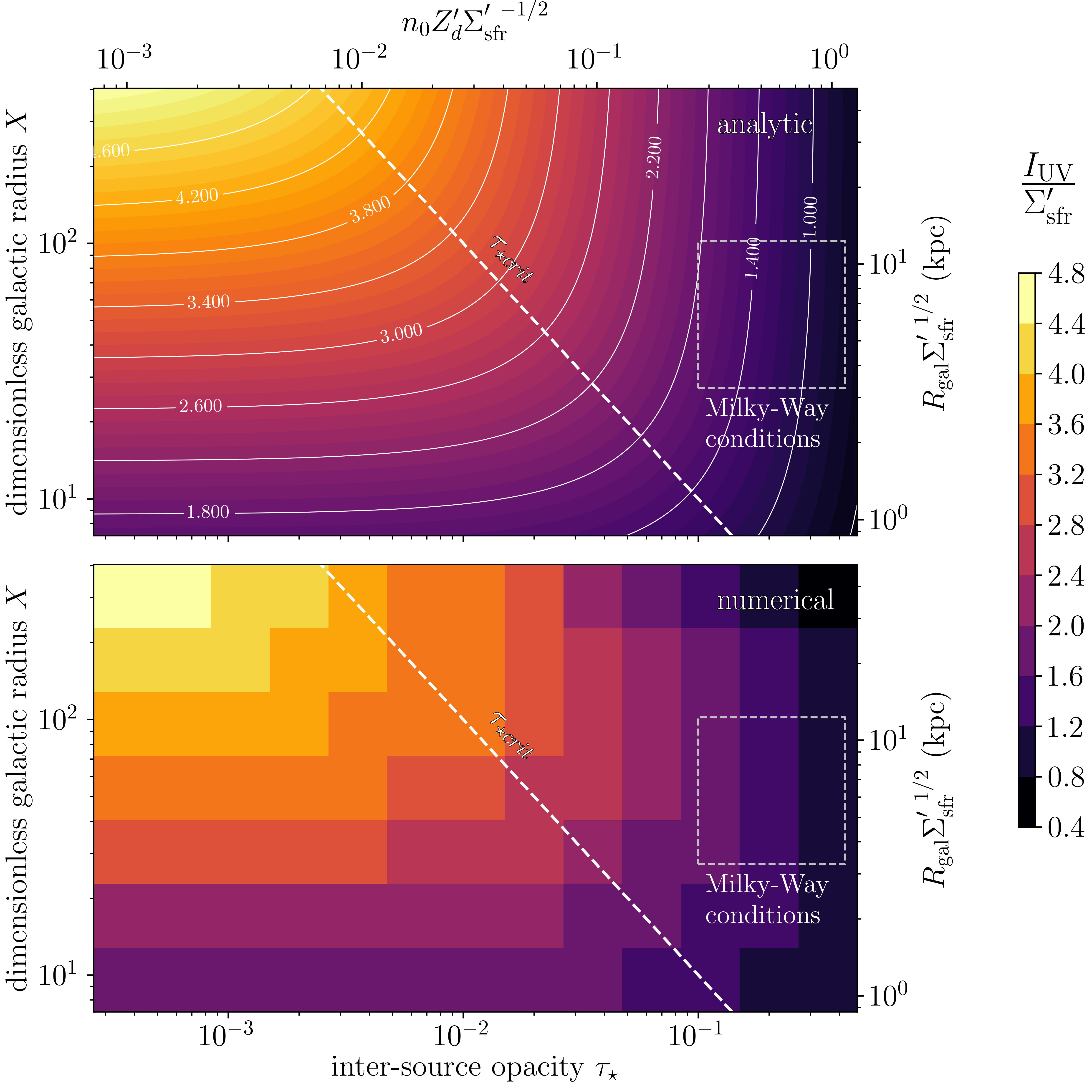} 
	\caption{
The FUV ISRF to the SFR density ratio (normalized to solar circle), 
$I_{\rm UV}/\Sigma_{\rm sfr}'$, in the $\tau_{\star}-X$ parameter space,  as given by the analytic model, Eq.~(\ref{eq: IUV/SFR}) (top) and as obtained by our numerical simulations (bottom), assuming a central observer ($\delta=0$).
On the $x$-axis, $\tau_{\star}$ is proportional to the DGR, 
and on the $y$-axis, $X$ is proportional to the galactic scale radius, as shown by the secondary axis (and Eq.~\ref{eq: l_star tau_star to physical}).
The dashed rectangle marks typical conditions for the Milky-Way disk.
The diagonal line is $\tau_{\rm \star crit}=1/X$ (Eq.~\ref{eq: tau_star_crit}, \ref{eq: Zd crit}) which separates ``the weak" (to the left) and ``the strong" (to the right) dust absorption regimes.
For $\tau_{\star}>\tau_{\rm \star crit}$ dust absorption controls the intensity of the ISRF. 
In this limit, $I_{\rm UV}/\Sigma_{\rm sfr}'$ increases with decreasing $\tau_{\star}$, 
and is independent of $X$.
In the opposite limit, the ISRF is controlled by the galaxy scale.
$I_{\rm UV}/\Sigma_{\rm sfr}'$ then increases with $X$ and is independent of $\tau_{\star}$.
}
		\label{fig: IUV_sfr_2D}
\end{figure*}

\begin{figure*}[t]
	\centering
	\includegraphics[width=1.0\textwidth]{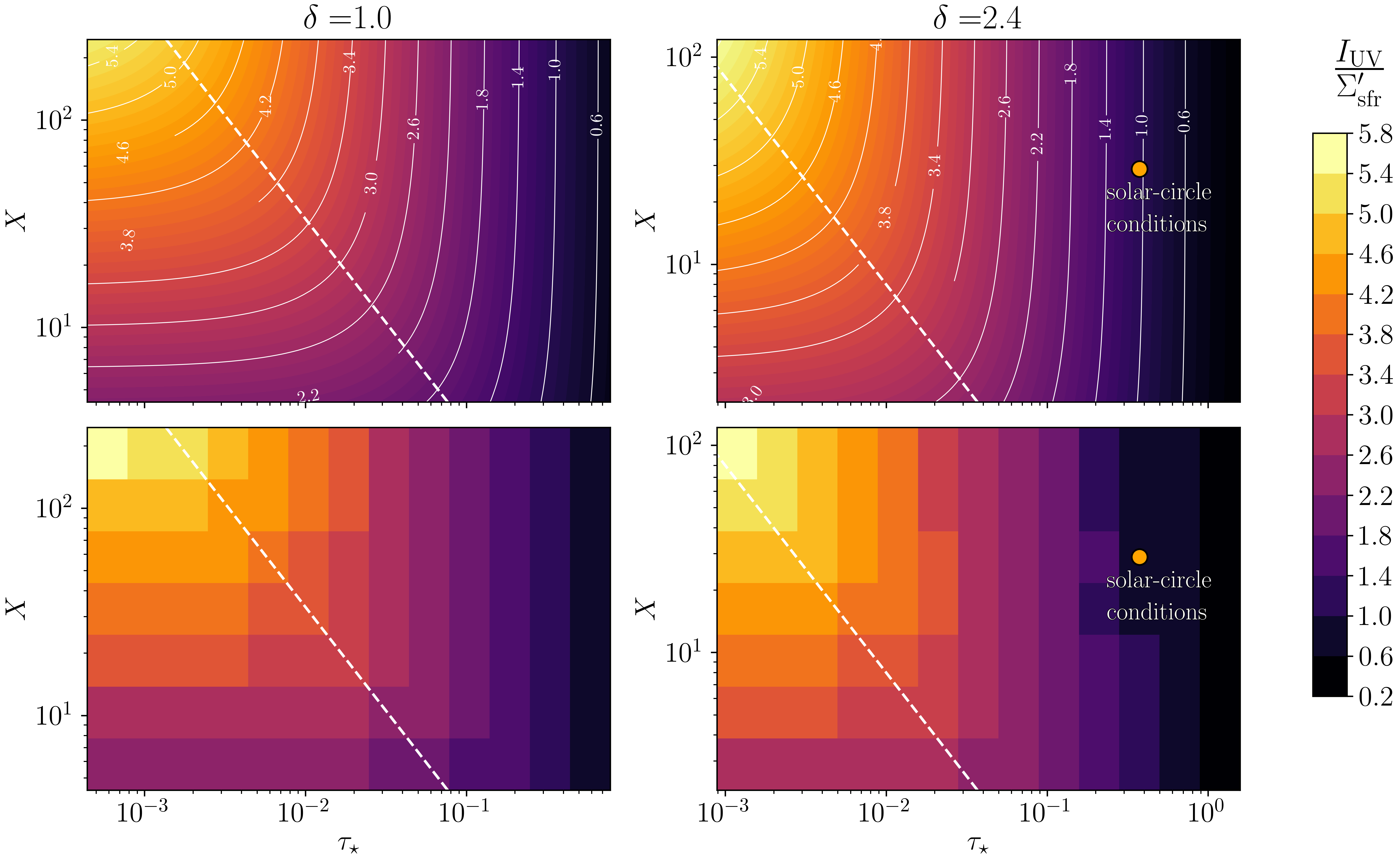} 
	\caption{
As Fig.~\ref{fig: IUV_sfr_2D} but for non-central observers, $\delta=R_{\rm obs}/R_{\rm gal}=1$ and 2.4.
The dashed line is the modified $\tau_{\rm \star crit}$ (Eq.~\ref{eq: tau_crit non center}). The solar circle conditions are indicated.
}
		\label{fig: IUV_sfr_2D_other_locations}
\end{figure*}

\subsection{The $I_{\rm UV}$ to star-formation rate ratio}
\label{sub: IUV/sfr}
We now rewrite the normalized flux in Eq.~(\ref{eq: f_th}) in physical units.
Defining $I_{\rm UV} \equiv F/F_0$ where $F_0 = 2.7 \times 10^{-3}$ erg cm$^{-2}$ s$^{-1}$ is the solar neighborhood ISRF \citep{Draine1978}, and with $\Sigma_{\star}=12 \ \Sigma_{\rm sfr}^{\prime} \ L_{\odot} \ {\rm pc}^{-2}$ (Eq.~\ref{eq: Sigma sfr}), we get  
\begin{equation}
\label{eq: IUV/SFR}
    \frac{I_{\rm UV}}{\Sigma_{\rm sfr}'} = 0.91 
    \left(
    \frac{\mathrm{e}^{-\tau_{\star}\bar{x}_0}}{2 \pi \bar{x}_0^2} 
    +E_1 \left[ 
    \bar{x}_0\tau_{\star}+\frac{\bar{x}_0}{X} 
    \right] 
\right) \ .
\end{equation}
where $\bar{x}_0=0.9$ and $\tau_{\star}$ and $X$ encapsulate the dependence on 
the DGR, SFR and galactic scale radius (Eq.~\ref{eq: l_star tau_star to physical}).
The $I_{\rm UV}-\Sigma_{\rm sfr}'$ ratio expresses how much FUV interstellar radiation flux  is gained per massive star formed. 

The $I_{\rm UV}-\Sigma_{\rm sfr}'$ ratio  as a function of $\tau_{\star}$ (at our fiducial $X=96$) is shown in Fig.~\ref{fig: IUV_vs_Zd}.
The solid and dashed curves are the median values as obtained from our numerical simulations, and our analytic model, Eq.~(\ref{eq: IUV/SFR}), respectively. 
They are in excellent agreement.
The dashed line shows $\tau_{\rm \star, crit}$ as given by Eq.~(\ref{eq: tau_star_crit}). 
To the right of this line we are in the strong dust absorption limit.
As expected, in this regime the median $I_{\rm UV}/\Sigma_{\rm sfr}'$ increases as $\tau_{\star}$ decreases, and as dust absorption becomes less efficient and allows larger regions of the galaxy disk to contribute to the ISRF.
As $\tau_{\star}$ decreases below $\tau_{\star  \rm crit}$, the galactic disk becomes optically thin and the ISRF saturates.

The shaded region in Fig.~\ref{fig: IUV_vs_Zd} is the IQR as obtained from our numerical simulations.
It is a measure of the dispersion in the intensity of the FUV ISRF.
These fluctuations in $I_{\rm UV}$ result from random fluctuations in the positions and luminosities of the FUV sources in respect to the observer, as captured by the different realizations of the FUV source distributions in our numerical models (see Fig.~\ref{fig: 2d sources} for an example of one such realization).
These fluctuations reflect the fact that (a) for a given observer position, the distribution of FUV sources changes as OB stars form and die with time (\citetalias{Parravano2003}), and (b) at a given instant of time, the distance to FUV sources changes for variations in the observer position.
The IQR spans $\Delta(I_{\rm UV}/\Sigma_{\rm sfr}') \approx 1.1$ at high $\tau_{\star}$, and $\approx 1.8$ at low $\tau_{\star}$.
However, when divided by the median value, the relative dispersion decreases with decreasing $\tau_{\star}$, from $1.1$ at high $\tau_{\star}$, to 0.46 at low $\tau_{\star}$.
This is because fluctuations in the ISRF are dominated by nearby FUV sources. 
As $\tau_{\star}$ decreases, far-away sources contribute more to the ISRF, and thus the relative dispersion in $I_{\rm  UV}$ decreases.

In Fig.~\ref{fig: IUV_sfr_2D} we show the $I_{\rm UV}/\Sigma_{\rm sfr}'$ ratio (median) in the 2-dimensional $\tau_{\star}-X$ parameter space, as given by our analytic model (Eq.~\ref{eq: IUV/SFR}) (top), and by our numerical simulations (bottom).
Over most of the parameter space the two agree to within 5\% (see upper-left panel of Fig.~\ref{fig: comparison}).
The dashed rectangle shows typical conditions for the Milky-Way.
The diagonal line is the critical line, $\tau_{\star \rm crit}=1/X$ that separates the parameter space into the weak and strong dust absorption limits.
As predicted by the analytic model, to the right of this line, the ISRF becomes independent of $X$ and
depends only on $\tau_{\star}$ (increasing with decreasing $\tau_{\star}$),
whereas to the left of $\tau_{\rm \star crit}$ 
the ISRF intensity depends only on $X$.
As shown by the top and right axis, $\tau_{\star}$ is proportional to the DGR, and $X$ is proportional to the galactic scale radius.
The  additional (weak) dependence of $\tau_{\star}$ and $X$ on the SFR reflects 
the fact that as the SFR increases, the typical distance to nearby source decreases (see \S \ref{sub: physical units}).
This in turn results in a weak increase of $I_{\rm UV}/\Sigma_{\rm sfr}'$ with increasing SFR density.



Eq.~(\ref{eq: IUV/SFR}) (and Fig.~\ref{fig: IUV_sfr_2D}) cannot be directly compared to the 
FUV ISRF in the solar neighbourhood because
of  the assumption of a central observer.
In contrast, in the Milky-Way, the solar galactocentric radius, $R_{\odot}=8.5$ kpc, is large compared to the OB galactic scale radius, $R_{\rm gal}=3.5$ kpc.
We shall now generalize Eq.~(\ref{eq: IUV/SFR}) to the case of a non-central observer.

\subsection{Non-central Observer}
\label{sub th non center}
When the observer is located off the disk center the polar symmetry is  broken and the problem becomes more complicated.
In analogy to Eq.~(\ref{eq: f_th}), we may write the ISRF as
 \begin{align}
 \label{eq: F/NL non central}
     \frac{F}{\Sigma_{\star}} = \frac{\mathrm{e}^{-\tau_{\star}\bar{x}_{0}}}{4 \pi \bar{x}_0^2} 
     +  \int_0^{2 \pi} \int_{\bar{x}_0}^{\infty} \frac{1}{4 \pi x} \mathrm{e}^{-\tau_{\star}x-g(\frac{x}{X}, \delta, \theta)} \mathrm{d}x \mathrm{d}\theta \ .
\end{align}
Here we defined 
\begin{equation}
    g \equiv \left[ \left(\frac{x}{X} \right)^2 +2\delta  \left(\frac{x}{X} \right) \cos\theta+\delta^2 \right]^{1/2} - \delta \ .
\end{equation}
Recall $\delta=R_{\rm obs}/R_{\rm gal}$ is the galactocentric radius of the observer relative to the galactic scale radius.
In the limit $\delta \ll 1$, the function $g \rightarrow x/X$ and Eq.~(\ref{eq: F/NL non central}) approaches Eq.~(\ref{eq: f_th}).
When $\delta$ is not small,  
the integral cannot be  solved analytically.
However, as $F/\Sigma_{\star}$ is normalized to the local density of sources, 
its variation with the observer position is mild.
This is because a significant fraction of the ISRF is always contributed by sources close to the observer.
In fact, when $\tau_{\star}$ is large such that $\tau_{\rm gal} \gg 1$, $F/\Sigma_{\star}$ becomes 
independent of the observer position (and of $X$), as the ISRF is coming exclusively from sources close to the observer (whose density is characterized by $\Sigma_{\star}$). Galactic gradients are then insignificant.
On the other hand, when $\tau_{\star}$ is small, sources at distances of order $R_{\rm gal}$ are contributing to the ISRF and $F/\Sigma_{\star}$ then depends on $\delta$ (and on $X$).

Motivated by this analysis, we seek a simple approximation for $F/\Sigma_{\star}$ (or, equivalently $I_{\rm UV}/\Sigma_{\rm sfr}'$) that 
obeys the above criteria.
By examining our  numerical  results,
we find that Eqs.~(\ref{eq: f_th}) and (\ref{eq: IUV/SFR}) may be used to describe off-center observers
under the replacement
\begin{align}
\label{eq: X_delta}
    \bar{x}_0 \rightarrow \bar{x}_\delta &\equiv \bar{x}_0 \  (1 - 0.1 \delta^{0.5}) \\ \nonumber
    X \rightarrow X_\delta &\equiv X \ (1+2\delta^2) \ . 
\end{align}
where $\bar{x}_0=0.9$ (Eq.~\ref{eq: x0 eff}).
This form has the desired behaviour discussed above:
1) For a central observer, $\bar{x}_\delta = \bar{x}_0$ and $X_\delta = X$, and Eqs.~(\ref{eq: f_th}, \ref{eq: IUV/SFR}) are unaltered.
2) In the strong-dust-absorption regime, at $Z_{d}' > Z_{d,{\rm crit}}'$, 
the ISRF remains insensitive to $\delta$.
This is because in Eq.~(\ref{eq: X_delta}), the strongest dependence on $\delta$ is in $X_\delta$.
However, in this regime the ISRF is in any case only a function of $\tau_{\star}$.
3) In the weak-dust-absorption limit,
the ISRF does vary with $\delta$, as non local sources are important contributors to the flux.
The fact that the ISRF {\it increases} with $\delta$, reflects the increased contribution of sources towards disk-center, where $\Sigma_{\star}$ is exponentially higher.

A direct comparison of the numerical and analytic results for the entire parameter space is presented in Fig.~\ref{fig: comparison}.
Note that in this plot the span of $\tau_{\star}$ and $X$ vary with $\delta$ (the hatched regions are regions not covered by our simulations) because our numerical galaxy models
are defined by the central galaxy quantities ($\tau_{\star c}$, $X_c$) (see \S \ref{sub: numerical procedure}).
In contrast  $\tau_{\star}$ and $X$ are the local
quantities (at the observer position), and they vary with $\delta$
 (see Eq.~\ref{eq: local to central}).
For most of the parameter space considered the analytic approximation and numerical results are in good agreement, on the level of $\lesssim 10\%$ relative difference.

In Fig.~\ref{fig: IUV_sfr_2D_other_locations} we show $I_{\rm UV}/\Sigma_{\rm sfr}'$
as as function of the $\tau_{\star}$ and $X$ (at the observer position),
as given by our analytic solution, Eq.~(\ref{eq: x0 eff}, \ref{eq: IUV/SFR}, \ref{eq: X_delta}) (upper panels), and as obtained by our numerical simulations (bottom panels), for observers located at $\delta=1$, and, 2.4.
In the right panels we mark the point that corresponds to the solar circle, for which $\delta=2.4$, $\tau_{\star}=0.37$ and $X=29$ (see \S \ref{sub: physical units}).
At this point $I_{\rm UV}/\Sigma_{\rm sfr}'=1$, by definition.
Comparing the results for $\delta=1$ and 2.4, as well as with the central observer case $\delta=0$ (Fig.~\ref{fig: IUV_sfr_2D}) we see that for high $\tau_{\star}$ and/or high $X$, the ratio $I_{\rm UV}/\Sigma_{\rm sfr}'$ remains weakly sensitive to $\delta$.
As discussed above, this is because in the strong dust absorption limit the ISRF is generated by nearby sources, and thus when normalized to the local emissivity $\Sigma_{\star} \propto \Sigma_{\rm sfr}'$, it becomes insensitive to the observer position.
However, when $\tau_{\star}$ is small and/or when $X$ is small, sources at distances on the order of the galactic scale contribute significantly to the ISRF buildup, and $I_{\rm UV}/\Sigma_{\rm sfr}$ is $\delta$-dependent.
For peripheral observers (e.g., at $\delta=2.4$), $I_{\rm UV}/\Sigma_{\rm sfr}$ 
increases above the values seen by a central observer.
For example, for $X=50$ and at the smallest $\tau_{\star}$, we see that $I_{\rm UV}/\Sigma_{\rm sfr}$ increases from $\approx 3.8$ at $\delta=0$
(Fig.~\ref{fig: IUV_sfr_2D}; bottom panel), 
to 4.4-5.2 at $\delta=1-2.4$ (Fig.~\ref{fig: IUV_sfr_2D_other_locations}; bottom panels).
This increase results from the contribution of faraway sources towards the galaxy center where the density of sources is high compared to that at the observer position.

A varying observer position also alters the location of the transition line from strong-to-weak dust absorption regimes.
For a central observer we had $\tau_{\rm \star crit}=1/X$.
 Similarly, for a non-center observer we have
 \begin{equation}
 \label{eq: tau_crit non center}
     \tau_{\rm \star, crit} = 1/X_{\delta} = (1+2 \delta^2)^{-1} X^{-1} \ ,
 \end{equation}
or in terms of the DGR
  \begin{equation}
   \label{eq: Z_crit non center}
     Z'_{d,{\rm crit}} =
     0.32 (n_0 R_{\rm gal, kpc})^{-1}  (1+2 \delta^2)^{-1} \ .
 \end{equation}
 Thus, $\tau_{\rm \star, crit}$ and $Z'_{d,{\rm crit}}$ decrease with $\delta$, and a larger region in the parameter space is occupied by the strong-dust absorption regime.
 The critical $\tau_{\rm \star, crit}$ as given by Eq.~(\ref{eq: tau_crit non center}) is shown by the diagonal dashed lines in Fig.~\ref{fig: IUV_sfr_2D_other_locations}.
 Eq.~(\ref{eq: tau_crit non center}) is in qualitative agreement with the numerical results where we see that with increasing $\delta$ the transition from the strong-to-weak dust absorption regime shifts down and to the left towards lower $\tau_{\star}$ and $X$ values.
 However, a quantitative comparison shows that Eqs.~(\ref{eq: tau_crit non center}-\ref{eq: Z_crit non center}) under-estimate $\tau_{\rm \star, crit}$ and $Z'_{d,{\rm crit}}$ at large $\delta$.

\section{Summary and Discussion}
\label{sec: discussion}

\subsection{Summary}
In this paper we have studied how the FUV
ISRF strength varies with galactic properties, the dust to gas ratio, the galaxy scale radius, the gas density, and the star-formation rate density.
Two basic dimensionless parameters encapsulate these dependencies:
\begin{enumerate}
    \item the opacity over the inter-source distance, $\tau_{\star}=\sigma n l_{\star}$
    \item and the dimensionless disk radius $X=R_{\rm gal}/l_{\star}$
\end{enumerate}
(Eq.~\ref{eq: l_star tau_star to physical}).
The ISRF intensity to the SFR density ratio is described by
\begin{equation}
    \frac{I_{\rm UV}}{\Sigma_{\rm sfr}'} = 0.91 
    \left(
    \frac{\mathrm{e}^{-\tau_{\star}\bar{x}_0}}{2 \pi \bar{x}_0^2} 
    +E_1 \left[ 
    \bar{x}_0\tau_{\star}+\frac{\bar{x}_0}{X} 
    \right] 
\right) \ \nonumber
\end{equation}
(Eq.~\ref{eq: IUV/SFR}), where the parameter $\bar{x}_0=0.9$ accounts for varying luminosity FUV sources, i.e., OB stellar associations.
Eq.~(\ref{eq: IUV/SFR}) also applies for off-center locations in the galaxy disk under the replacement $\bar{x}_0 \rightarrow 0.9(1-0.1\delta)$ and $X\rightarrow X(1+2\delta^2)$ (Eq.~\ref{eq: X_delta}) where
$\delta=R_{\rm obs}/R_{\rm gal}$ is the ratio of the galactocentric radius of the observer to the galactic scale radius of the OB stars.
Over most of the parameter space considered, Eq.~(\ref{eq: IUV/SFR}, \ref{eq: X_delta}) is accurate to the level of $10$\%.

There are two basic regimes in the problem, the strong and weak dust-absorption regimes, separated by the critical line $\tau_{\rm \star, crit}=1/X$ or the critical dust-to-gas ratio $Z_{\rm d,  crit}$ (Eq.~\ref{eq: tau_star_crit}, \ref{eq: Zd crit}, or Eqs.~\ref{eq: tau_crit non center}, \ref{eq: Z_crit non center} for $\delta>0$).
In the weak-dust absorption regime ($Z'<Z_{\rm d,  crit}'$), the limit applicable to low metallicity dwarf and high redshift galaxies, sources on galactic scales contribute to the ISRF and $I_{\rm UV}/\Sigma_{\rm sfr}'$ is determined solely by the $X$ parameter.
In this limit the ISRF intensity increases with the galactic scale radius.
On the other hand, in the strong-dust absorption limit ($Z'>Z_{\rm d,  crit}$), i.e., the limit applicable to the Milky-Way,
the ISRF is limited by dust absorption.
In this limit $I_{\rm UV}/\Sigma_{\rm sfr}'$ is determined solely by the $\tau_{\star}$ parameter, and $I_{\rm UV}/\Sigma_{\rm sfr}'$ then increases with decreasing $\tau_{\star}$ (or with decreasing DGR or gas density).


\subsection{Regulation of Star-Formation}
\label{sub: SF regulation}
Star formation in galactic disks may be self regulated by a natural feedback process that occurs in the multiphase ISM.
Theories and observations suggest that the neutral ISM 
is dominated by two phases, the diffuse-warm ($T \sim 10^4$ K) and dense-cold ($\sim 100$ K) neutral media (WNM and CNM), where the CNM-WNM are in rough pressure equilibrium with a density contrast of $\approx 100$ \citep[e.g.,][]{Field1969, Heiles2003, Wolfire2003, Murray2018, Bialy2019}. 
These phases are steady-states at which the cooling and heating rates are equal.
The feedback loop operates as follows:
If the SFR of a galaxy (or a region in a galaxy) increases significantly, the intensity of the FUV ISRF also increases and results in excessive gas heating (through the photoelectric heating process) which results in the removal of the cold phase.
Since star-formation is efficient in cold and dense gas, this in turn leads to a reduction in the star-formation rate.
This feedback loop has been at the base of the star-formation regulation theory of \citet[][]{Parravano1988, Parravano1989} and \citet{Ostriker2010}.

A key ingredient in the star-formation self-regulation theory is
how much gain in the FUV ISRF is produced per SFR density,  i.e., what is the $I_{\rm UV}$-to-$\Sigma_{\rm sfr}'$ ratio, where the $I_{\rm UV}$ parameter determines the thermal pressure of the multiphase neutral ISM \citep{Wolfire1995, Bialy2019}.
Following \citet{Ostriker2010},  this thermal pressure (with additional pressure from turbulent motions) must balance the galactic disk weight which is determined by the stellar and dark matter density and gas surface density of the disk.
\citet{Ostriker2010} assumed a constant ratio for $I_{\rm UV}/\Sigma_{\rm sfr}'$ independent of galactic properties (see their Eq.~16).
Our finding that the $I_{\rm UV}/\Sigma_{\rm sfr}'$ ratio increases towards low 
metallicities implies that for the galaxy to still maintain a thermal pressure that allows a multiphase ISM, the SFR density must be reduced.
This means that low metallicity galaxies are naturally pushed to have a lower star-formation efficiency (SFR density divided by gas column density), compared to their high metallicity counterparts, i.e., a lower normalization for the Kennicutt-Schmidt relation \citep[e.g.,][]{Schmidt1959, Kennicutt1998, Bigiel2008}.

More recent studies \citep{Kim2011, Ostriker2011, Shetty2012, Kim2013b, Kim2015, Kim2017} focused on the role of turbulent pressure, generated by SN momentum injection as the dominant SFR regulation mechanism.
These studies find that for a given SFR density, the turbulent pressure surpasses the thermal pressure, 
and thus suggest that SN feedback controls the SFR.
Our finding that the $I_{\rm UV}$-to-$\Sigma_{\rm sfr}'$ ratio increases in low metallicity galaxies implies that thermal pressure may dominate over turbulent pressure potentially reviving the role of FUV heating of the multiphase ISM as the dominant regulator of star-formation in these galaxies.
This opens up a window for a bimodality in the star-formation process in galaxies, where in high metallicity galaxies it is controlled by SN feedback,  whereas in low metallicity galaxies it is controlled by FUV heating of the multiphase ISM.
We note that at sufficiently low dust abundances, yet another transition occurs, 
as photoelectric heating becomes inefficient and other heating mechanisms become 
dominant: X-ray, cosmic-ray, and H$_2$-formation heating \citep{Bialy2019}.


\subsection{The Advantage of Using Dimensionless Quantities}
\label{sub: advantage of dimensionless form}
Identifying the dimensionless parameters and rewriting the problem (and solving it) in dimensionless form (i.e., $F/\Sigma_{\star}$ as a function of $\tau_{\star}$ and $X$; Eqs.~\ref{eq: f_basic}, \ref{eq: f_th}) has a big advantage.
It makes the results more general and enables future modifications to any assumed values and/or scaling relations to physical parameters that enter the model.
For example, if one wishes to revise the value of the dust absorption cross section
(here assumed to be $\sigma_g=10^{-21}$ cm$^2$), all one needs to do is to re-scale the relation between $\tau_{\star}$ and $Z'_d$ accordingly (Eq.~\ref{eq: l_star tau_star to physical}).
The numerical and analytic results for $f$ in terms of $\tau_{\star}$ are still valid and unchanged, it is only the translation from $\tau_{\star}$ to $Z_d'$ that changes.
Similarly, any variations to our assumed relation for the source emissivity and SFR (Eq.~\ref{eq: Sigma sfr}), do not affect our results for the dimensionless flux, $f$, as $f$ only depend on the dimensionless quantities, $\tau_{\star}$ and $X$.
Such a variation will of course affect the  flux in physical units, $F/\Sigma_{\rm sfr}'$ (or $I_{\rm UV}/\Sigma_{\rm sfr}'$).
If one wishes to adopt a different normalization (or a different scaling relation) than the one we used, one has to simply 
re-scale the $I_{\rm UV}/\Sigma_{\rm sfr}'$ that we computed, accordingly (see for example, \S \ref{subsub: IMF} below).

\subsection{Limitations and Future Model Extensions}
To be able to derive analytic formula for the variation of the ISRF with galactic properties we are forced to make simplifying assumptions.
Here we discuss these assumptions and their validity.

\subsubsection{Variations to the Initial Mass Function}
\label{subsub: IMF}
To relate the source emissivity, $\Sigma_{\star}$, to the SFR surface density, $\Sigma_{\rm sfr}'$, we have assumed that the two are proportional with a proportionality factor calibrated based on solar neighborhood conditions (Eq.~\ref{eq: Sigma sfr}).
If the stellar initial mass function (IMF) varies between galaxies, or within a galaxy, it would imply a variation in the adopted proportionality factor, as the sources that contribute to the ISRF are only the massive stars (mainly OB stellar types) which represent the high mass end of the IMF.
For example, for a top heavy IMF, the fraction of massive stars is high compared to standard IMF \citep[e.g.][]{Kroupa2002}. This will imply a higher $\Sigma_{\star}/\Sigma_{\rm sfr}'$ than adopted here.
Whether the IMF is constant or varies with galactic properties is still under debate, and thus we do not include IMF variations as a component in our model.
For a given IMF model it is possible to re-scale our results by adopting a non constant $\Sigma_{\star}-\Sigma_{\rm sfr}'$ proportionality factor that accounts for any potential systematic trend of the IMF with galactic properties, e.g., with metallicity (see \S \ref{sub: advantage of dimensionless form}).

\subsubsection{Finite Disk Scale Height}
In our models we have assumed thin disk geometry.
This is a valid approximation if the radius over which a substantial contribution  
to the ISRF is accumulated, is large compared to the galactic scale height.
We define $r_{1/2}$ as the radius within which half of the ISRF is accumulated.
Exploring our numerical results we find that for large $\tau_{\star}$, $r_{1/2} \approx 2 l_{\star}$ (e.g., see Fig.~\ref{fig: cummulative}). 
For low $\tau_{\star}$ the total ISRF is higher as the contribution of distant sources is more substantial. 
In this case we find $r_{1/2} \approx 5 l_{\star}$. 
Plugging the expression for $l_{\star}$ (Eq.~\ref{eq: l_star tau_star to physical}) and 
comparing $r_{1/2}$ to the scale height we obtain the requirement:
\begin{equation}
\label{eq: H requiremenet}
    H_{\rm OB} \ll H_{\rm OB, crit} \equiv 420 \Sigma_{\rm sfr}'^{-1/2} \phi' \ {\rm pc} \ ,
\end{equation}
where $H_{\rm OB}$ is the scale height of OB stars, and where we defined $\phi'=(r_{1/2}/l_{\star})/3.5$. 
The factor $\phi'$ is close to unity, it varies from $\approx 1.4$ at low $\tau_{\star}$ down to $\approx 0.6$ at high  $\tau_{\star}$.
Note that since $H_{\rm OB}$ is scale height of {\it only} the OB (massive) stars, it is much smaller than the stellar disk height which includes all stellar populations.

For typical condition Eq.~(\ref{eq: H requiremenet}) is satisfied.
For example, in the Milky Way \citet{Wolfire2003} finds $H_{\rm OB}=59$ pc in the solar circle and towards smaller galactocentric radii, and an increasing trend towards larger galactocentric radii.
In contrast, $\Sigma_{\rm sfr}'$ varies from $\approx 3.6$ to $\approx 0.066$ for radii $R=4$ kpc to 18 kpc, implying  $H_{\rm OB, crit} =220-1600$ pc.
For very active starbursts, or very puffy disks, requirement (\ref{eq: H requiremenet}) may be violated and a treatment of the disk width is needed.
Another situation that requires considering the finite thickness of the disk is for studies of fluctuations of the ISRF distribution, in particular the high end of the ISRF
distribution.
This is because, as discussed in \citetalias{Parravano2003} the high end of the ISRF distribution often results from situations where the distance between the observer and the nearby source is abnormally small.
For these cases $r_{1/2}$ will be smaller than the median $r_{1/2}$ value discussed above, and may be smaller than $H_{\rm OB}$.
We generalize the theory to the case of finite disk thickness and study fluctuations in the FUV ISRF in a forthcoming study.

\section{Conclusions}
\label{sec: conclusions}
We have modeled the  FUV interstellar radiation field both numerically and analytically.
The ratio of the FUV ISRF intensity to the SFR density depends on the DGR (or metallicity), gas density, galactic scale radius and the SFR density. These dependencies are encapsulated in two controlling dimensionless parameters, $\tau_{\star}$ and $X$.
The $\tau_{\star}-X$ parameter space is separated into two basic regimes, the weak and the strong dust absorption regimes, separated by the critical line $\tau_{\rm \star crit}=1/X$, where
for $\tau_{\star}>\tau_{\rm \star crit}$ the ISRF is limited by dust absorption, and for $\tau_{\star}<\tau_{\rm \star crit}$ it is limited by the galaxy scale radius.
With decreasing $\tau_{\star}$ (or decreasing metallicity), the ISRF intensity per SFR density increases, and reaches values a factor of 3-6 higher compared to disks like the Milky-Way.
This may have important implications on the thermal state of the neutral ISM and the star-formation efficiency in low metallicity galaxies (e.g., dwarfs and high redshift galaxies), and may introduce a natural feedback loop for star-formation.

\acknowledgements
SB thanks David Hollenbach and Chris McKee for insightful discussions and useful comments.
SB acknowledges support from the Harvard-Smithsonian Institute for Theory and Computation (ITC).

\end{document}